\documentclass[10pt,journal,twocolumn]{IEEEtran}

\usepackage{times}
\usepackage{graphicx}
\usepackage{array}
\usepackage{booktabs}
\usepackage{multirow}
\usepackage{url}
\usepackage{amsmath}
\usepackage{cite}
\usepackage{xcolor}

\title{On a General Theoretical Framework for Radio Frequency Fingerprint-Based Authentication}

\author{Yuanyu~Zhang,~\IEEEmembership{Member, IEEE},~Jianing Wang,~Shuangrui~Zhao,~\IEEEmembership{Member, IEEE},~Pinchang~Zhang,~\IEEEmembership{Member, IEEE},~Yulong~Shen,~\IEEEmembership{Senior Member, IEEE},~and~Xiaohong~Jiang,~\IEEEmembership{Senior Member, IEEE}%
\thanks{Corresponding author: Yulong Shen.}
\thanks{Y.~Zhang, J.~Wang, S.~Zhao and Y.~Shen are with the School of Computer Science \& Technology, Xidian University, Xi'an, Shaanxi 710071, China, and also with Shaanxi Key Laboratory of Network and System Security. Email: \{yyuzhang, zhaoshuangrui\}@xidian.edu.cn, ylshen@mail.xidian.edu.cn}
\thanks{P.~Zhang is with the College of Cybersecurity, Tarim University, Alaer, Xijiang 843300, China. Email: 120260003@taru.edu.cn}
\thanks{X.~Jiang is with the School of Systems Information Science, Future University Hakodate, Hakodate 041-8655, Japan. Email: jiang@fun.ac.jp.}
}

\begin{document}
\maketitle

\begin{abstract}
While radio frequency fingerprint (RFF)-based wireless device authentication has been widely studied across different datasets and scenarios, there still lacks a fundamental theory to explain why and how RFF can serve as a reliable device identity, significantly hindering the practical application of such an authentication technology. In this article, we integrate the RFF modeling with authentication property analysis to propose a general theoretical framework to facilitate the development of such a theory. The RFF modeling process reveals how RFFs are induced, evolved and observed along the transmitter-channel-receiver chain, built upon which, the authentication property analysis process then outlines how the trustworthiness of an RFF should be examined in terms of its uniqueness, stability, distinguishability, and unforgeability. By linking the RFF formation/evolution to these authentication properties, the framework offers a solid foundation for understanding why and how RFF-based authentication is trustworthy in practice. We also discuss the communication-authentication co-design issue based on the theoretical insights from the proposed framework.
\end{abstract}


\section{Introduction}
Radio frequency fingerprint (RFF)-based authentication exploits hardware imperfections in wireless transmitters to infer the source of a received signal \cite{jqzhang2025tifs}. Because such imperfections are naturally embedded in wireless waveforms, RFFs have attracted broad interest for lightweight authentication, spoofing detection, and signal-source verification in Internet of Things (IoT) \cite{jqzhang2023commag}, unmanned aerial vehicle (UAV) \cite{zxcai2025tifs}, vehicle-to-everything (V2X) \cite{xyqin2024tifs}, 5G/6G cellular \cite{hyluo2025tifs}, and satellite networks \cite{gabriela2023tifs}.

The research community has made impressive progress in RFF-based authentication. Classical approaches extract handcrafted features such as carrier frequency offset \cite{gxshen2021jsac}, phase noise \cite{khatab2025machine}, I/Q imbalance \cite{xlgu2026tmc}, or modulation errors \cite{jshe2024twc}. More recent studies have employed deep learning to learn discriminative representations directly from I/Q samples, spectrograms, constellation images, or other signal views \cite{gxshen2023commag,jnwang2026tifs,srzhao2026tifs}. These studies have improved device identification accuracy and robustness under many experimental settings.

However, existing RFF-based authentication studies remain largely empirical. They often focus on feature extraction, classifier design, robustness enhancement, and performance evaluation in specific datasets or scenarios. High accuracy in these settings does not by itself explain why RFFs can serve as reliable evidence of device identity, nor does it clarify when such evidence remains trustworthy in practical deployment. A learned model may capture transmitter-intrinsic hardware features, but it may also exploit environment-specific patterns, receiver-dependent artifacts, signal-configuration biases, or other shortcuts that are not genuine source evidence. The lack of a fundamental and systematic theory has therefore become a key barrier to the practical application of RFF-based authentication.

This article takes a step toward addressing this gap by proposing a general framework that supports the development of such a theory. The framework integrates RFF modeling with authentication property analysis. RFF modeling examines how RFFs are induced, evolved, and observed along the transmitter-channel-receiver chain. Authentication property analysis studies the key properties that determine whether the observed RFFs can support trustworthy source verification, including uniqueness, stability, distinguishability, and unforgeability. By linking RFF formation/evolution to authentication property analysis, the framework provides a structured basis for understanding why and when RFF-based authentication can be reliable.

The contributions of this article are threefold. First, we identify the absence of a fundamental and systematic theory as a key bottleneck in RFF-based authentication research and propose a general framework for developing the theoretical foundations of RFF-based authentication. Second, we introduce a cascaded formation and evolution model of RFFs, which distinguishes transmitter-side intrinsic RFFs from receiver-side observed RFFs and clarifies how hardware-induced signal features become practical authentication evidence. Third, we organize the foundational analysis around four key authentication properties, namely uniqueness, stability, distinguishability, and unforgeability, and discuss how the resulting theoretical insights can support communication-authentication co-design.

This article does not aim to benchmark a specific RFF-based authentication algorithm or dataset. Instead, it provides a conceptual framework for identifying what should be modeled, analyzed, and justified before RFFs can be trusted as authentication evidence. To the best of our knowledge, this is one of the first attempts to systematically study RFF-based authentication foundations by connecting RFF modeling with authentication property analysis.

\section{Framework for RFF-based Authentication Theory Development}
\begin{figure}[!t]
\includegraphics[width=\linewidth]{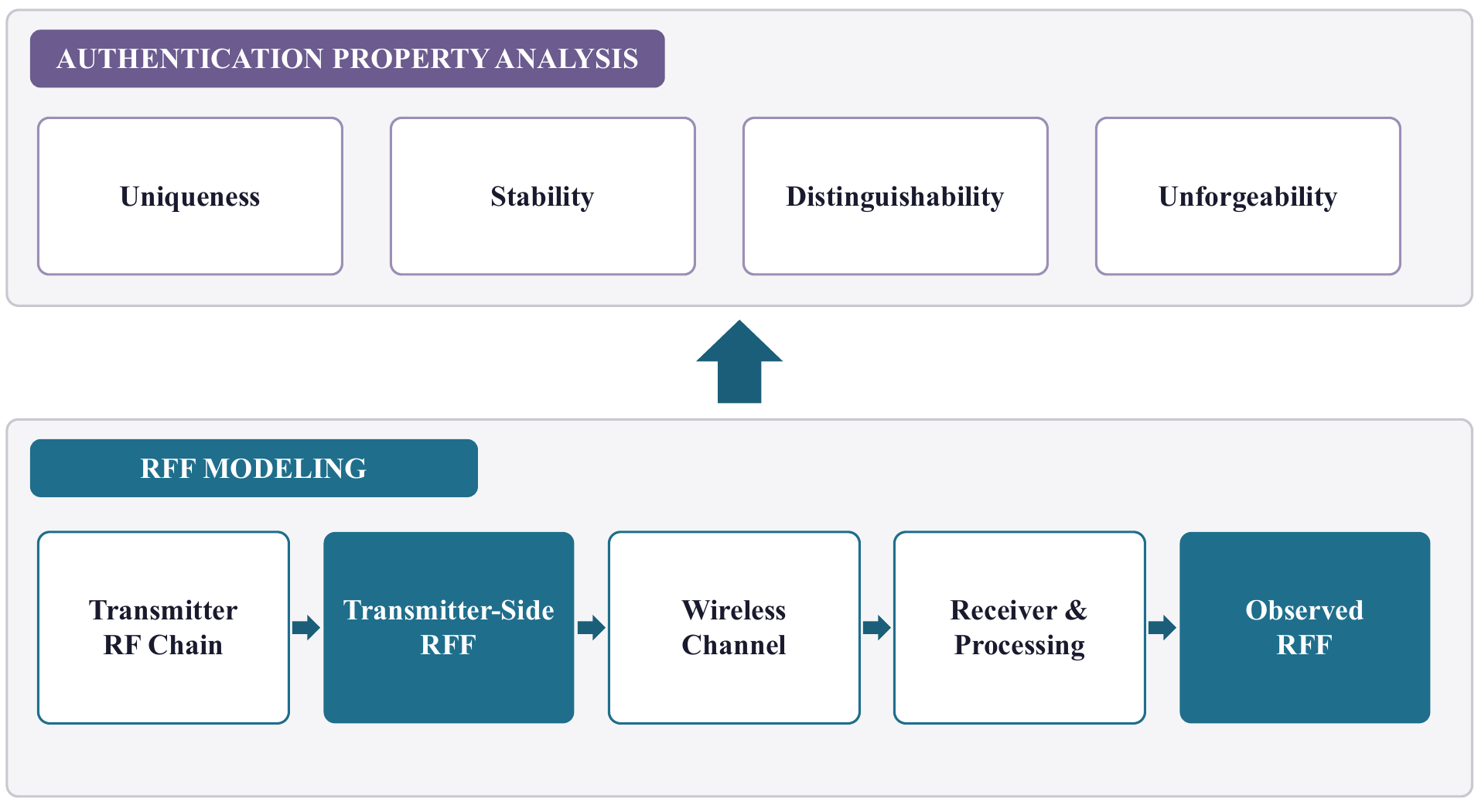}
\caption{A framework for RFF-based authentication theory development.}
\label{fig:framework}
\end{figure}

This section presents the proposed framework. As illustrated in Fig.~\ref{fig:framework}, the framework integrates RFF modeling with authentication property analysis. These two parts jointly address how RFFs are formed and observed, and what properties should be examined to determine whether such RFFs can support source authentication.

RFF modeling provides the basis for understanding the origin and observation of RFF evidence. At the transmitter side, hardware-induced signal features are generated and reshaped by RF components such as oscillators, mixers, I/Q modulators, power amplifiers, filters, and antennas. These features are then carried by the wireless signal through the propagation channel and further affected by fading, noise, interference, mobility, and environmental dynamics. At the receiver side, receiver hardware, synchronization, sampling, compensation, and feature extraction procedures transform the received signal into the observable RFFs used by the authentication system. This modeling view connects the physical origin of RFFs with the practical observations available to a receiver.

Authentication property analysis provides guidance for studying four key properties of RFFs: uniqueness, stability, distinguishability, and unforgeability. Uniqueness asks whether different devices possess physically different RFFs. Stability asks whether RFFs remain sufficiently consistent under practical variations. Distinguishability asks whether different devices can be reliably separated in the receiver-side observation domain. Unforgeability asks whether an attacker can stably generate forged signals whose observed RFFs pass authentication. These properties are not isolated metrics; they specify different prerequisites for using RFFs as authentication evidence.
 
Overall, the framework follows a bottom-up logic. RFF modeling explains where RFF evidence comes from and how it becomes observable. Authentication property analysis specifies what should be examined based on the RFF model. Communication-authentication co-design is not treated as an additional part of the framework; instead, it is discussed later as a system-level design implication supported by the framework. In this way, the framework focuses on the foundations of RFF-based authentication while still providing a basis for guiding practical communication-authentication design.

\section{RFF Modeling}

\begin{figure*}[t]
\centering
\includegraphics[width=\textwidth]{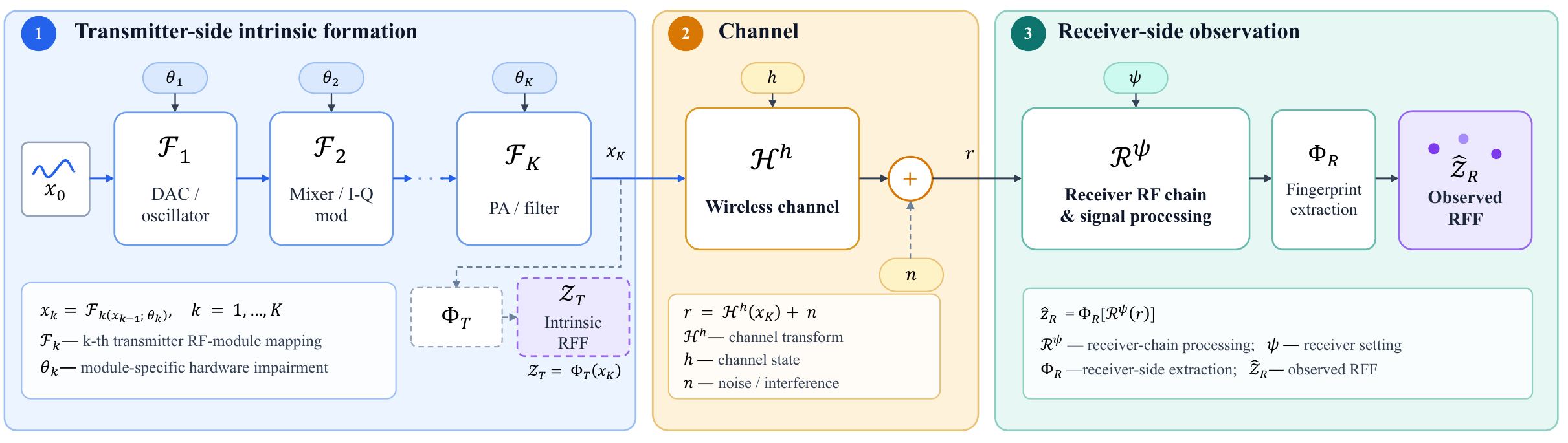}
\caption{Cascaded formation and evolution model of RFFs.}
\label{fig:rff-cascaded-formation-model}
\end{figure*}

RFF modeling provides the modeling basis for the proposed framework. It is not meant to prescribe a universal RFF model for all RFF-based authentication systems. Instead, it specifies what should be made explicit before authentication properties are analyzed: how transmitter-side hardware-induced signal features are formed, how they are transformed by the channel, and how the receiver obtains the observed RFFs used for decisions. In this view, an RFF is not a static feature attached to a transmitter, but the observable outcome of a cascaded formation and evolution process, as illustrated in Fig.~\ref{fig:rff-cascaded-formation-model}. Because different studies may target different signal formats, hardware platforms, frequency bands, receiver architectures, and threat scenarios, the model should make its assumptions explicit rather than hiding them inside a classifier or dataset. 

\subsection{Transmitter-Side Intrinsic RFF Formation}

At the transmitter side, RFFs originate from hardware nonidealities in components such as digital-to-analog converters, oscillators, mixers, I/Q modulators, power amplifiers, filters, and matching networks. Manufacturing tolerances, circuit mismatch, thermal effects, aging, calibration residuals, and nonlinearities introduce hardware-induced signal features into the emitted waveform. These features may appear as carrier-frequency offset, phase noise, I/Q imbalance, nonlinear distortion, transient behavior, modulation errors, or other device-dependent effects.

These features should be understood as the result of a cascaded RF-chain process rather than isolated component effects. Each module may introduce new features while reshaping those inherited from upstream modules through amplification, suppression, coupling, or filtering. For example, power-amplifier nonlinearity may interact with I/Q imbalance, and filtering may suppress some distortion components while preserving others. This is why the modeling part of the framework emphasizes cascaded formation and evolution rather than a single impairment source. It also prevents the analysis from attributing an RFF to one visible feature while ignoring how downstream modules may change, mask, or enhance that feature.

Let the transmitter RF chain contain $K$ modules. The input signal is denoted by $x_0$, and the output of the $k$-th module is denoted by $x_k$. The $k$-th module is characterized by a transformation function $\mathcal{F}_k(\cdot)$ and hardware impairment parameters $\theta_k$. The cascaded transmitter-side signal formation can be conceptually written as
$x_K =\left(\mathcal{F}_K^{\theta_K} \circ \mathcal{F}_{K-1}^{\theta_{K-1}} \circ \cdots \circ \mathcal{F}_1^{\theta_1}\right)(x_0)$.
Here, $x_K$ is the transmitted waveform carrying accumulated hardware-induced signal features; it is not itself the RFF. To separate the signal carrier from the RFF evidence, we denote the transmitter-side intrinsic RFF as $z_T=\Phi_T(x_K)$, where $\Phi_T(\cdot)$ is an analysis-oriented mapping that extracts transmitter-specific features after suppressing task-dependent factors such as payload, modulation, bandwidth, power, and packet structure. This notation does not assume that $z_T$ is directly observable; it clarifies the physical source on which uniqueness analysis should be based.

\subsection{Channel-Induced Transformation}

After transmitter-side formation, the RFF-bearing signal propagates through the wireless channel. Fading, delay spread, Doppler shift, interference, mobility, and environmental dynamics may alter the observability of transmitter-side features. Some components may be masked by channel variations, while others may remain robust under certain propagation conditions. Hence, physical uniqueness at the transmitter does not automatically imply receiver-side distinguishability.

The channel should therefore be explicitly included in the RFF model. Otherwise, an RFF-based authentication model may appear to distinguish devices because the dataset contains label-correlated channel conditions rather than genuine transmitter evidence. Depending on the target system, channel modeling may involve path loss, multipath, interference, mobility, synchronization error, or domain shifts across location and time. The key is not necessarily to remove the channel completely, but to clarify how transmitter-dependent evidence and propagation-induced variations are coupled. This distinction is especially important for cross-domain evaluation: a feature that appears stable in one location or receiver setting may actually encode the environment rather than transmitter identity.

\subsection{Receiver-Side RFF Observation}

The receiver does not directly observe the intrinsic RFF. It observes a distorted signal processed by its own RF front end and signal-processing chain, including automatic gain control, synchronization, sampling, filtering, compensation, channel estimation, equalization, and feature extraction. Therefore, the RFFs used by an authentication system are receiver-side and observation-dependent.

Let $\mathcal{H}^{h}(\cdot)$ denote the channel transformation with channel condition $h$, $n$ denote noise and interference, $\mathcal{R}^{\psi}(\cdot)$ denote receiver-side hardware and front-end processing with receiver parameters $\psi$, and $\Phi_R(\cdot)$ denote the receiver-side RFF extraction function. The observed RFF can be conceptually described as
$
\hat{z}_R =
\Phi_R \left[
\mathcal{R}^{\psi}
\left(
\mathcal{H}^{h}(x_K) + n
\right)
\right].
%\label{eq:observed-fingerprint}
$
This expression highlights that $\hat{z}_R$ is jointly shaped by transmitter hardware, the wireless channel, receiver hardware, and signal processing. The intrinsic RFF explains where source evidence comes from, while the observed RFF determines what evidence is actually available for authentication.

This distinction also defines the analysis objects for authentication property analysis. Uniqueness is mainly rooted in intrinsic transmitter-side sources, whereas stability, distinguishability, and unforgeability must be analyzed in the receiver-side observation domain. Without this model, high classification accuracy may reflect channel shortcuts, receiver artifacts, or signal-format biases rather than genuine source evidence. RFF modeling therefore connects the physical origin of RFFs with their practical observations and provides the basis for property-level analysis. In later sections, the model is not treated as an end in itself; it is used to specify which RFF space, which observation process, and which uncertainty sources each foundational property should consider.

\section{Authentication Property Analysis}
Built upon RFF modeling, authentication property analysis identifies the properties that should be examined before RFFs are treated as trustworthy source evidence. These properties are not simple performance metrics. They specify the questions that researchers should answer, using the RFF model defined above, when explaining the mechanism of RFF-based authentication.

We organize this analysis around four properties: uniqueness, stability, distinguishability, and unforgeability. Uniqueness concerns the physical origin of RFF evidence; stability concerns whether such evidence remains consistent under practical variations; distinguishability concerns whether different devices can be separated in the receiver-side observation domain; and unforgeability concerns whether an attacker can stably imitate legitimate RFF evidence.

\subsection{Uniqueness}
Uniqueness is the physical prerequisite of RFF-based authentication. It asks whether different devices possess physically different RFF sources. In the proposed framework, this property is mainly rooted in the transmitter-side intrinsic RFF, which reflects hardware-induced signal features before channel and receiver transformations. These features may originate from manufacturing tolerances, circuit mismatch, nonlinearities, oscillator imperfections, I/Q imbalance, power-amplifier characteristics, filters, antennas, and other RF-chain components.

However, uniqueness should not be assumed simply because hardware imperfections are random. Random variations do not automatically imply that the resulting RFF representations are sufficiently diverse, measurable, or collision-free. If two transmitters can produce the same, or practically indistinguishable, intrinsic RFF representation, no receiver-side classifier can establish source identity in a principled way. Thus, uniqueness should be analyzed through RFF-space capacity, population-level dispersion, and collision risk, rather than inferred only from finite-dataset separability.

A concrete route is to abstract the intrinsic RFF of each fixed device, under fixed or normalized operating conditions, as a high-dimensional vector. Across a population of independently manufactured devices, these vectors collectively induce a statistical distribution in the intrinsic RFF space. Uniqueness can then be studied by asking how dispersed this population-level distribution is and how likely two devices are to become indistinguishable under a given quantization precision, measurement resolution, or tolerance requirement.

\begin{figure}[!t]
\centering
\includegraphics[width=\linewidth]{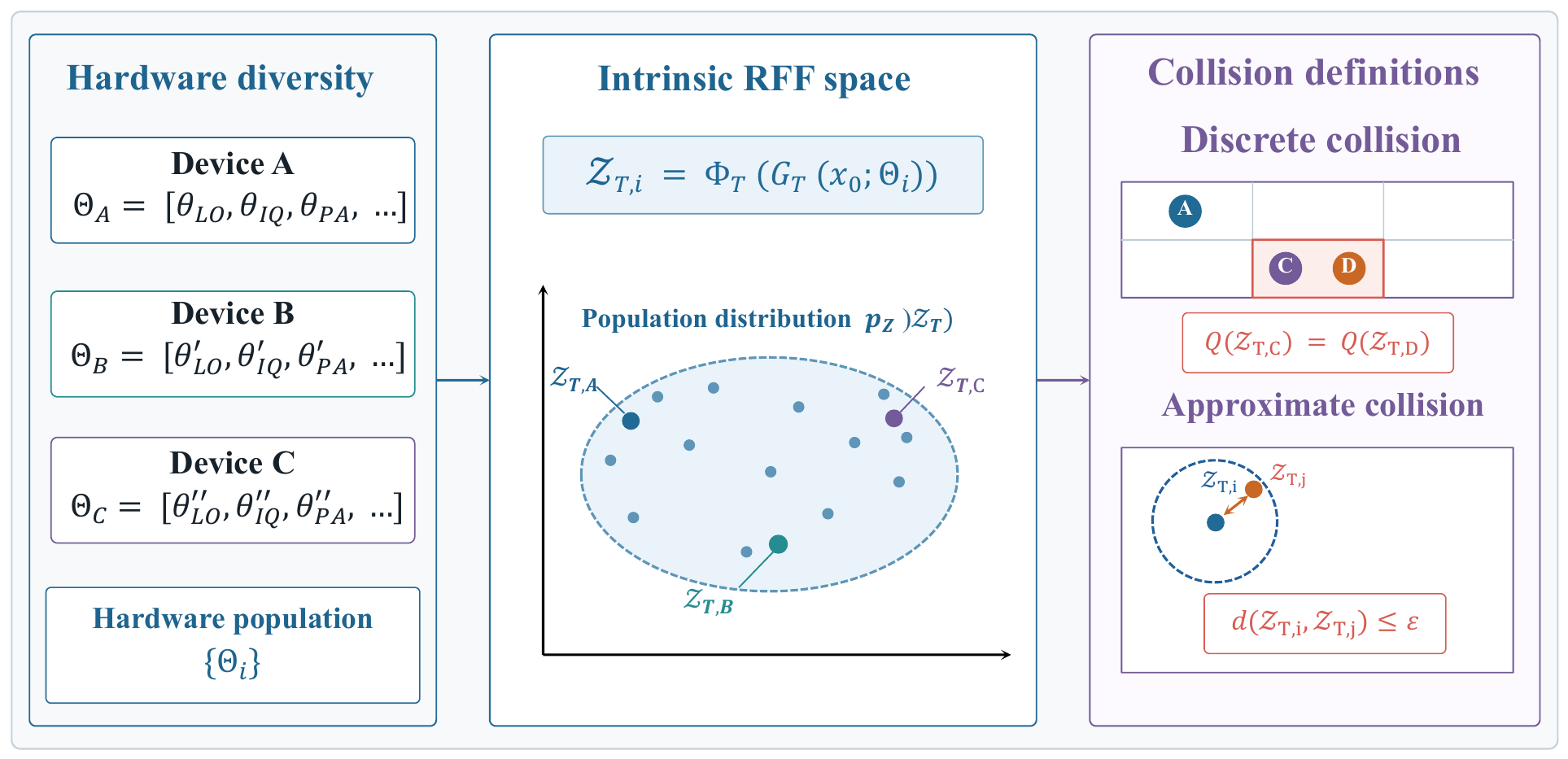}
\caption{Collision-oriented view of RFF uniqueness in the intrinsic RFF space.}
\label{fig:rff_uniqueness_intrinsic_collision}
\end{figure}

This leads naturally to collision-oriented analysis, as illustrated in Fig.~\ref{fig:rff_uniqueness_intrinsic_collision}. If an intrinsic RFF vector is mapped into a finite code space, a collision occurs when two devices are assigned to the same codeword or quantization cell. In a continuous space, an approximate collision may be defined by a distance below the resolution required by the authentication system. Collision probability and collision entropy provide one possible way to quantify effective RFF diversity. They are not the only metrics, but they help move uniqueness analysis beyond qualitative statements that hardware imperfections are ``naturally different''. Other routes may examine sensitivity to hardware-parameter variations, the minimum distance among intrinsic RFF points, or the probability that two devices become indistinguishable under realistic measurement constraints. The common principle is that uniqueness should be analyzed in the intrinsic RFF space defined by the model, not merely inferred from a particular dataset.

\subsection{Stability and distinguishability}
Stability and distinguishability are closely coupled because both are evaluated in the receiver-side observation domain. Stability asks whether the observed RFFs of the same device remain sufficiently consistent under practical variations, while distinguishability asks whether the observed RFFs of different devices remain reliably separable. Geometrically, stability concerns the compactness and drift of each device-specific observation region, whereas distinguishability concerns the separation among such regions.

Neither property alone is sufficient. A representation may be stable but not distinctive if different devices form compact yet overlapping regions. It may also appear distinctive under controlled conditions but become unstable across channels, receivers, time, SNR, modulation formats, or operating states. The key question is therefore whether observation regions remain both bounded and separated under the variations expected in deployment.

A useful analysis route is to represent each device by a distribution in the observed RFF space. This distribution is shaped by transmitter hardware, channel conditions, receiver hardware, synchronization, compensation, and feature extraction. Stability can be studied through intra-device spread, temporal drift, and domain shift; distinguishability can be studied through inter-device distance, margin, overlap probability, or open-set separability. Reliable authentication requires the intra-device variation to remain smaller than the inter-device separation under practical conditions. This view also connects naturally to threshold design: the acceptance region must be wide enough to tolerate legitimate variation, but narrow enough to avoid overlap with other devices and impostors.

\begin{figure}[!t]
\centering
\includegraphics[width=\linewidth]{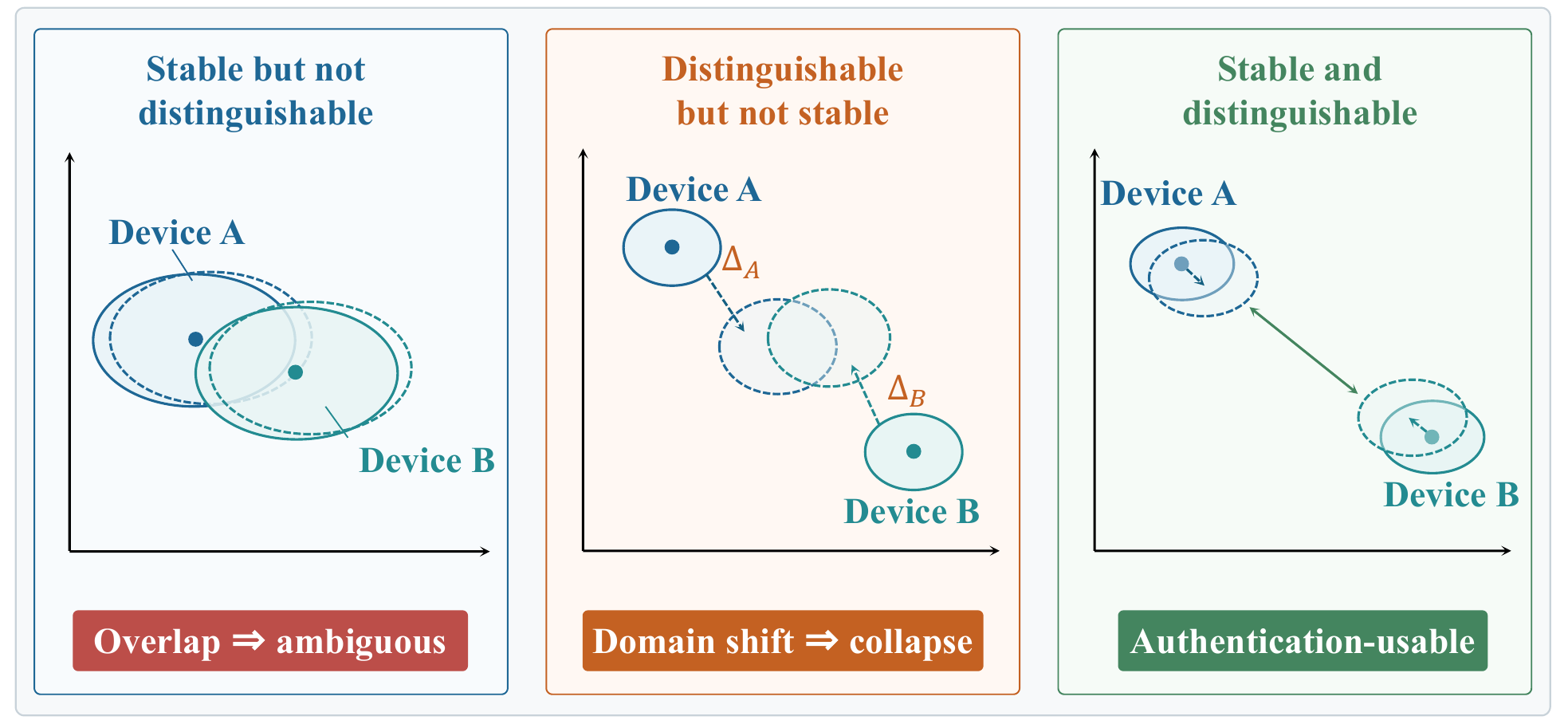}
\caption{Observation-region view of RFF stability and distinguishability.}
\label{fig:rff_stability_distinguishability_observation}
\end{figure}

This observation-region view also explains why uniqueness does not automatically imply practical decision evidence. Two devices may have different intrinsic RFFs, but their observed distributions may overlap after channel and receiver transformations, as shown in Fig.~\ref{fig:rff_stability_distinguishability_observation}. Conversely, a classifier may separate devices by exploiting receiver or environment artifacts. Stability and distinguishability should therefore be examined under cross-domain conditions and with realistic negative samples, rather than only through closed-set classification accuracy. In practice, stability and distinguishability may need to be assessed at multiple time scales, from packet-level observations to session-level decisions. Longer observations or repeated measurements may improve separation, but they also introduce latency and resource cost, which motivates the later discussion on communication-authentication co-design.

\subsection{Unforgeability}
Unforgeability is the security prerequisite of RFF-based authentication. It asks whether an attacker can generate signals whose observed RFFs are accepted as belonging to a legitimate device. In this framework, unforgeability is defined in the receiver-side observation domain: the attacker does not need to reproduce the legitimate transmitter internally, but only needs to make the target receiver observe a forged RFF inside the legitimate acceptance region.

The difficulty of forgery comes from a long and nonlinear transformation chain. An attacker may need to observe legitimate signals, estimate RFF-relevant features, synthesize a forged waveform, transmit it through its own imperfect RF chain, pass through the wireless channel, and survive the target receiver's processing. Each step can introduce estimation, modeling, waveform-generation, hardware-control, channel-compensation, or receiver-mismatch errors. These errors may be coupled, transformed, and amplified rather than simply added. This makes RFF forgery different from reproducing a digital identifier: the attacker must control an analog transformation chain whose output is judged only after channel and receiver effects.

\begin{figure*}[t]
\centering
\includegraphics[width=0.8\textwidth]{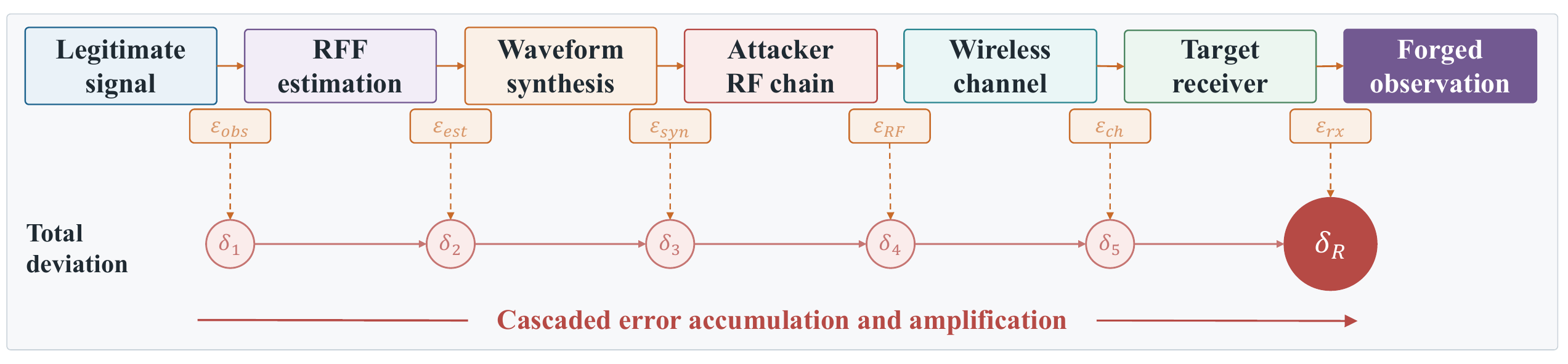}
\caption{A cascaded error-propagation view of RFF forgery.}
\label{fig:rff_unforgeability}
\end{figure*}

This motivates a cascaded error-propagation view of RFF forgery, as illustrated in Fig.~\ref{fig:rff_unforgeability}. A small estimation mismatch may be reshaped by the attacker's RF front end; waveform-generation error may be amplified by nonlinearities or mixed with I/Q imbalance; channel-compensation error or receiver-side processing may further move the forged observation away from the intended target. Thus, forgery should not be modeled as one-shot waveform similarity.

A more appropriate abstraction is stable hitting in the observation domain. The attacker succeeds only if forged signals can repeatedly and controllably produce observed RFFs inside the legitimate acceptance region across packets, time instants, channels, and receiver states. This view suggests that unforgeability should be analyzed through the complete attacker-side chain, the propagation of errors, and the probability that the final forged observation remains inside the legitimate region. Freshness checks, temporal consistency, channel context, and challenge-response mechanisms can make stable hitting more difficult. This analysis also shows why unforgeability should be tied to an explicit attacker model. Replay attackers, same-model hardware impersonators, software-defined radio attackers, and adaptive attackers with receiver feedback pose different levels of risk and should not be collapsed into a single benign negative class.

\section{Communication-Authentication Co-Design}

\begin{figure*}[t]
\centering
\includegraphics[width=0.8\textwidth]{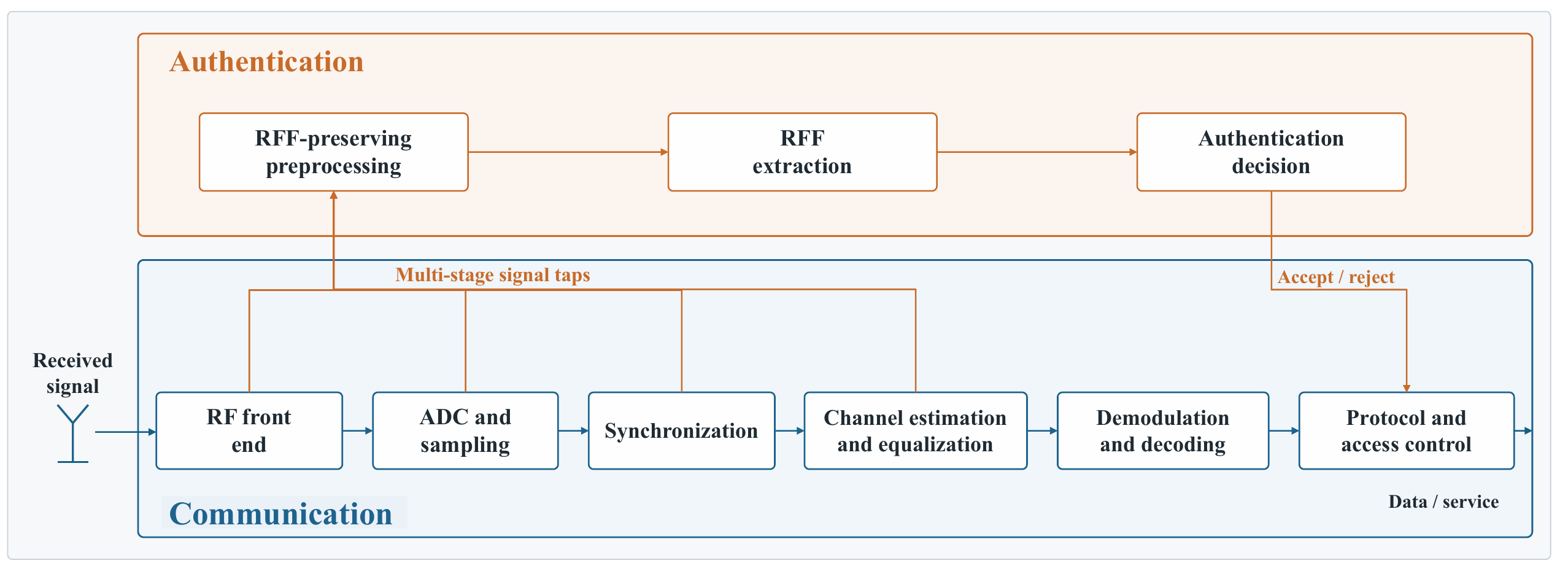}
\caption{Illustration of Communication-authentication co-design.}
\label{fig:rff_communication_authentication_codesign}
\end{figure*}

This section discusses communication-authentication co-design as an important design implication supported by the framework. The basic idea is that once the origin, observation process, and authentication properties of RFFs are clarified, they can guide practical wireless-system design. RFF-based authentication should therefore not be treated as an external classifier attached to a receiver, but should be considered together with the communication mechanisms that generate, preserve, transform, and consume RFF evidence, as illustrated in Fig.~\ref{fig:rff_communication_authentication_codesign}.

Communication and authentication are coupled because they share the same signal, hardware, and processing chain. Communication systems use synchronization, frequency-offset compensation, I/Q calibration, channel estimation, equalization, automatic gain control, power control, coding, modulation, and resource allocation to improve reliability and efficiency. These operations are usually optimized for bit error rate, throughput, latency, energy consumption, and coverage, but they may also reshape the evidence used for authentication. A processing step that improves demodulation may suppress a hardware-induced feature; a compensation algorithm that removes distortion may also remove identity evidence; and a resource-allocation strategy that improves throughput may reduce the observation time or bandwidth available for authentication.

Authentication requirements also constrain communication design. Reliable RFF-based authentication may require sufficient observation length, sampling bandwidth, stable receiver processing, adaptive thresholds, freshness verification, and resistance to replay or forgery. These requirements consume resources, affect latency, and may limit aggressive compensation. They also interact with the properties in the previous section: compensation may improve stability but reduce uniqueness or distinguishability; longer observations may improve confidence but increase delay; tighter thresholds may improve unforgeability but increase false rejection. Thus, communication choices can change the geometry of observed RFF regions, and authentication requirements can reshape the communication design space. The goal of co-design is not to sacrifice communication performance for authentication, but to determine where identity evidence should be preserved, where it should be normalized, and how much resource should be spent to obtain a trustworthy decision.

\subsection{A Constrained Optimization View}

The co-design problem can be understood as a constrained optimization problem in which communication performance, authentication reliability, security risk, and resource cost must be balanced simultaneously. Design choices such as waveform configuration, receiver processing, compensation strategy, RFF extraction, authentication threshold, observation length, and resource allocation affect both communication objectives and authentication objectives. For example, a receiver-processing or compensation strategy that improves demodulation may suppress identity-bearing RFF traces, while a longer observation window or more complex feature extractor may improve authentication confidence at the cost of latency, computation, energy consumption, or spectrum efficiency.

This view does not prescribe a universal objective function. Instead, it provides an organizational principle for RFF-based communication-authentication co-design. A practical design should maintain communication quality, preserve authentication properties such as stability, distinguishability, and resistance to forgery, and keep resource consumption within acceptable limits. Different systems may instantiate these requirements according to their signal formats, hardware platforms, receiver architectures, application constraints, and threat models. In this sense, the proposed framework helps designers clarify which communication utility is being protected, which authentication property is being strengthened, and which resource or security constraint limits the overall design.

\subsection{Co-Design Principles}

A first principle is to decide which impairments should be compensated and which should be preserved. Conventional receivers treat hardware impairments as nuisance factors, yet some impairments are the source of RFF evidence. The goal is not to preserve all distortions, but to distinguish communication-harmful yet authentication-useful features from distortions that are unstable, receiver-dependent, or easy to forge.

A second principle is to make RFF extraction aware of communication context. The observed RFFs may depend on modulation, bandwidth, power, packet length, coding, pilot structure, channel condition, receiver configuration, and mobility. An RFF-based authentication system should either normalize these factors, condition its decision on them, or design signal resources that make RFF evidence more comparable across conditions.

A third principle is to jointly allocate authentication evidence and communication resources. Richer observations, such as longer packets, wider bandwidth, repeated transmissions, or multi-antenna measurements, may improve authentication confidence but increase latency, computation, energy, or spectrum cost. Observation length, decision thresholds, and verification frequency should therefore adapt to channel quality, risk level, device role, and application requirements.

A fourth principle is to incorporate security context into the communication protocol. RFF separability alone cannot guarantee unforgeability. Replay or adaptive forgery may succeed if freshness, timing, channel consistency, or protocol context is ignored. Challenge-response signaling, random pilots, temporal consistency checks, multi-packet verification, and cross-layer access control can help turn physical-layer evidence into stronger authentication decisions. These principles should be applied together rather than independently, because preserving a trace, adapting a threshold, or changing a pilot structure may simultaneously affect communication quality, stability, distinguishability, and forgery resistance.

\subsection{From Property Analysis to System Design}

The proposed framework supports co-design by translating modeling and property-level understanding into system-level design considerations. If RFF modeling identifies where identity-bearing features are generated and how they are transformed, receiver processing can be designed to preserve or expose useful evidence rather than blindly suppress it. If uniqueness analysis identifies diverse intrinsic sources, the system can choose signal formats and features that reveal these sources. If stability analysis reveals dominant sources of intra-device drift, compensation or context-aware modeling can be targeted more precisely. If distinguishability analysis shows where observation regions overlap, resources, features, or thresholds can be adjusted to enlarge decision margins. If unforgeability analysis identifies attacker-controllable paths, protocol mechanisms can be introduced to reduce replay opportunities, limit feedback leakage, and make stable hitting more difficult.

This view suggests that future RFF-based authentication systems should be evaluated beyond isolated identification accuracy. A practical system should report how RFF evidence is modeled, how stable and distinguishable it remains under communication variations, how difficult it is to forge under explicit attacker capabilities, and how authentication requirements affect resources, latency, and reliability. Communication-authentication co-design therefore reframes RFF-based authentication from a standalone recognition problem into a system-level trust problem: the objective is to jointly support reliable data transmission and trustworthy source verification under realistic resource, channel, hardware, and security constraints.

\section{Conclusion}
Radio frequency fingerprint (RFF)-based authentication has shown great promise, but its practical deployment requires a complete and fundamental theory. This article highlights that RFFs are not static features; their formation, evolution, and observation involve complex interactions among transmitter hardware, wireless channels, receiver processing, and environmental variations. It also shows that linking fingerprint models to uniqueness, stability, distinguishability, and unforgeability remains a challenging and highly coupled problem. Developing a fundamental and systematic theory for RFF-based authentication therefore requires sustained community efforts. We hope this article encourages further research on RFF-based authentication theory, paving the way for the widespread application of RFF-based authentication technology.

\section*{Acknowledgment}
This work was supported in part by the National Key R\&D Program of China (Grant No.  2023YFB3107500), in part by the National Natural Science Foundation of China (No. 62220106004, 92467201), in part by the Fundamental Research Funds for the Central Universities (Grant No. QTZX25080, ZDRC2202), in part by Shaanxi Elite Talent Introduction Program (Youth Project), in part by the Young Talent Fund of Association for Science and Technology in Shaanxi, China, in part by the Natural Science Basic Research Program of Shaanxi Province (2025JC-YBQN-869), in part by the Open Project of Shandong Provincial Key Laboratory of Independent and Reliable Computing Technology and Equipment (No.KT25500400-lab), in part by the Key R\&D Program of Shandong Province of China (No.2025CXPT089).

\bibliographystyle{IEEEtran}
\bibliography{reference}

\section*{Biographies}
\textbf{Yuanyu~Zhang} is with Xidian University. His research interests include radio frequency fingerprinting, GNSS security and Satellite Internet security.

\textbf{Jianing~Wang} is with Xidian University. Her research interests include radio frequency fingerprinting and physical-layer authentication.

\textbf{Shuangrui~Zhao} is with Xidian University. His research interests include communication systems, physical-layer security, and trustworthy wireless networks. 

\textbf{Pinchang~Zhang} is with Tarim University. His research interests include physical-layer authentication and unmanned systems security. 

\textbf{Yulong~Shen} is with Xidian University. His research interests include intelligent unmanned systems security, data security and AI for security. 

\textbf{Xiaohong~Jiang} is with Future University Hakodate. His research interests include wireless communications security covering physical-layer authentication, covert communication and secure communication. 

\end{document}